\documentclass[reprint,prl,aps]{revtex4-1}
\usepackage{graphicx}
\usepackage{color}
\usepackage{dcolumn}
\usepackage{bm}
\usepackage{amssymb}
\usepackage{braket}
\usepackage{color,amsmath}
\usepackage{gensymb}
\usepackage{soul,xcolor} 
\setstcolor{red} 
\usepackage{epstopdf}
\usepackage{hhline}
\usepackage{tabularx}

\newcolumntype{C}[1]{>{\centering\arraybackslash}m{#1}}
\newcolumntype{N}{@{}m{0pt}@{}}

\definecolor{cadmiumgreen}{rgb}{0.0, 0.42, 0.24}

\begin{document}

\title{Competing correlated states and abundant orbital magnetism in twisted monolayer-bilayer graphene}

\author{Minhao He$^{1}$}
\author{Ya-Hui Zhang$^{2}$}
\author{Yuhao Li$^{1}$}
\author{Zaiyao Fei$^{1}$}
\author{Kenji Watanabe$^{3}$} 
\author{Takashi Taniguchi$^{4}$} 
\author{Xiaodong Xu$^{1,5\dagger}$}
\author{Matthew Yankowitz$^{1,5\dagger}$}

\affiliation{$^{1}$Department of Physics, University of Washington, Seattle, Washington, 98195, USA}
\affiliation{$^{2}$Department of Physics, Harvard University, Cambridge, MA, USA}
\affiliation{$^{3}$Research Center for Functional Materials,
National Institute for Materials Science, 1-1 Namiki, Tsukuba 305-0044, Japan}
\affiliation{$^{4}$International Center for Materials Nanoarchitectonics,
National Institute for Materials Science,  1-1 Namiki, Tsukuba 305-0044, Japan}
\affiliation{$^{5}$Department of Materials Science and Engineering, University of Washington, Seattle, Washington, 98195, USA}
\affiliation{$^{\dagger}$ xuxd@uw.edu (X.X.); myank@uw.edu (M.Y.)}

\maketitle

\textbf{Flat band moir\'e superlattices have recently emerged as unique platforms for investigating the interplay between strong electronic correlations, nontrivial band topology, and multiple isospin `flavor' symmetries~\cite{Balents2020,Andrei2020}. Twisted monolayer-bilayer graphene (tMBG) is an especially rich system owing to its low crystal symmetry and the tunability of its bandwidth and topology with an external electric field~\cite{Ma2020,Park2020theory,Rademaker2020,Chen2020tMBG,Polshyn2020,Shi2020}. Here, we find that orbital magnetism is abundant within the correlated phase diagram of tMBG, giving rise to the anomalous Hall effect (AHE) in correlated metallic states nearby most odd integer fillings of the flat conduction band, as well as correlated Chern insulator states stabilized in an external magnetic field. The behavior of the states at zero field appears to be inconsistent with simple spin and valley polarization for the specific range of twist angles we investigate, and instead may plausibly result from an intervalley coherent (IVC) state with an order parameter that breaks time reversal symmetry. The application of a magnetic field further tunes the competition between correlated states, in some cases driving first-order topological phase transitions. Our results underscore the rich interplay between closely competing correlated ground states in tMBG, with possible implications for probing exotic IVC ordering.}

In twisted graphene heterostructures with flat electronic bands, Coulomb interactions can spontaneously lift the degeneracy between spin, orbital, and lattice flavor symmetries~\cite{Cao2018a,Cao2018b,Yankowitz2019,Lu2019,Balents2020,Andrei2020}. In the simplest case, the many-body ground state is completely polarized into a subset of these isospin flavors. However, a much wider family of correlated ground states are also possible, including various density wave orders~\cite{Isobe2018} and exotic quantum spin liquid states~\cite{Zhang2020}. Among these, theoretical calculations often find IVC states to be competitive with Ising-like valley polarized (VP) states at zero magnetic field~\cite{Zhang2019,Bultninck2020,Bultninck2020b}. In magic-angle twisted bilayer graphene (tBLG), certain IVC states have been proposed as the parent ground state out of which superconductivity emerges~\cite{You2019,Bultninck2020,Khalaf2020}, although direct experimental identification of the ground state order is challenging. Twisted monolayer-bilayer graphene (tMBG) features lower crystal symmetry, and consequentially the lattice degeneracy is strongly lifted at the single-particle level in an external displacement field, $D$. The lowest moir\'e conduction band has four remaining degenerate copies corresponding to spin and valley, and is flat enough to host correlated states over a small range of twist angles~\cite{Chen2020tMBG,Polshyn2020,Shi2020}. The bandwidth, $W$, and valley Chern number, $C_v$, additionally depend on the orientation of $D$ (Supplementary Information Fig.~\ref{fig:dispersion})~\cite{Ma2020,Park2020theory,Rademaker2020}, which can polarize charge carriers more strongly towards either the monolayer or Bernal-stacked bilayer graphene sheet~\cite{Chen2020tMBG}. The high tunability of the bands with the combination of twist angle, doping, $D$, and magnetic field makes tMBG an attractive platform for investigating the nature of closely competing correlated and topological ground states.

\begin{figure*}[t]
\includegraphics[width=6.9 in]{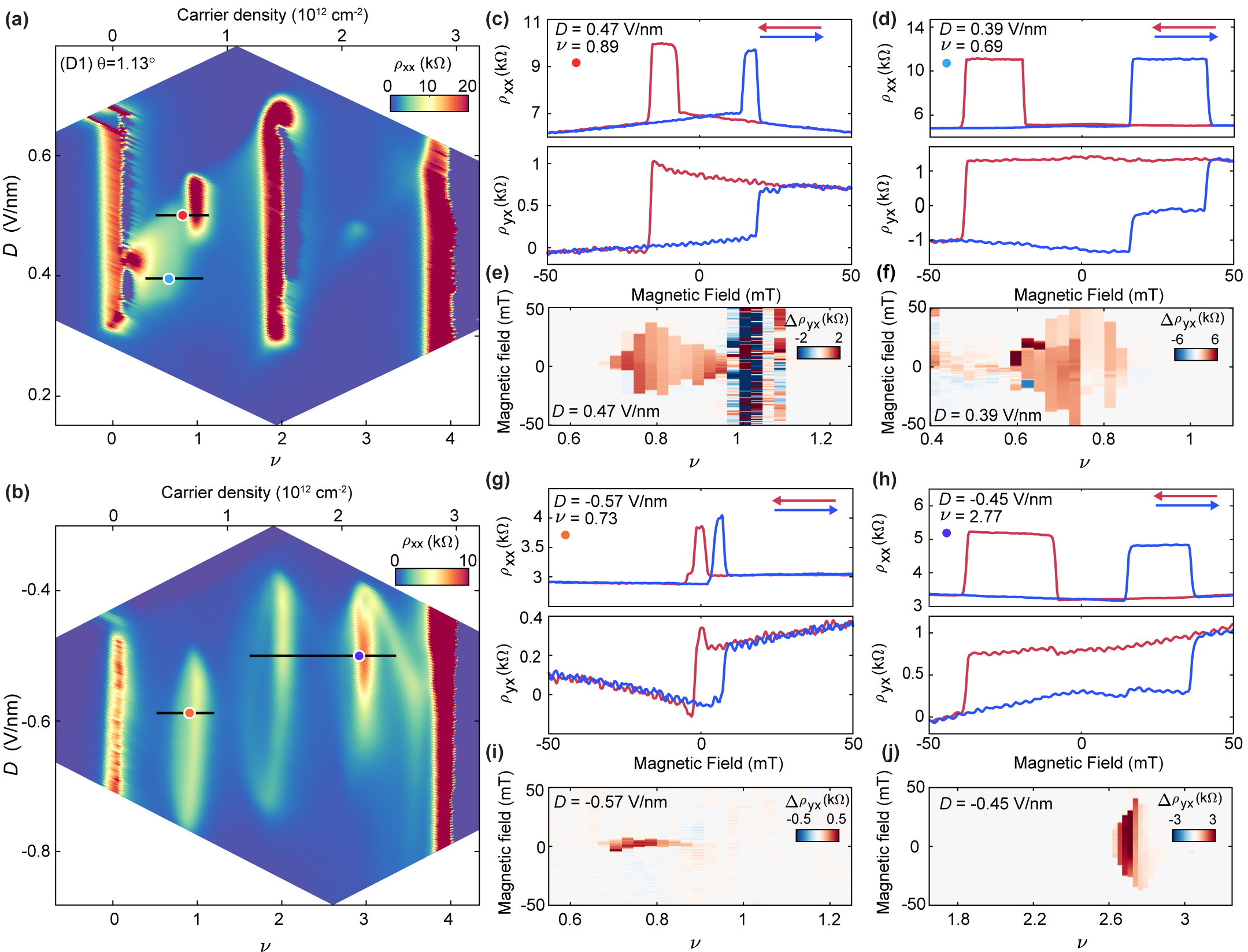} 
\caption{\textbf{AHE in metallic states of tMBG.}
\textbf{a-b}, Longitudinal resistivity, $\rho_{xx}$, of device D1 ($\theta=1.13^{\circ}$) for $D>0$ (\textbf{a}) and $D<0$ (\textbf{b}). $\rho_{xx}$ is symmetrized at $|B|=0.5$~T in order to suppress any magnetic hysteresis effects, but looks very similar at $B=0$ (Supplementary Information Fig.~\ref{fig:deviceD1}a).
\textbf{c-d}, $\rho_{xx}$ (top) and $\rho_{yx}$ (bottom) acquired as $B$ is swept back and forth at $\nu$ and $D$ indicated by the labels, and by the red and blue markers in \textbf{a}, respectively.
\textbf{e-f}, Hysteresis loop height, $\Delta \rho_{yx}$, as a function of doping at $D=0.47$~V/nm (\textbf{e}) and $D=0.39$~V/nm (\textbf{f}), as indicated by the black lines in \textbf{a}. Rapidly oscillating red and blue points near $\nu=1$ in \textbf{e} arise due to the correlated trivial insulating state rather than magnetic ordering (see Supplementary Information Fig.~\ref{fig:nu1_absentAHE}a-b).
\textbf{g-h}, $\rho_{xx}$ (top) and $\rho_{yx}$ (bottom) acquired as $B$ is swept back and forth at $\nu$ and $D$ indicated by the labels, and by the orange and purple markers in \textbf{b}, respectively.
\textbf{i-j}, $\Delta \rho_{yx}$ as a function of doping at $D=-0.57$~V/nm (\textbf{i}) and $D=-0.45$~V/nm (\textbf{j}), as indicated by the black lines in \textbf{b}.
We note that small oscillation features in $\rho_{yx}$ curves are noise, arising due to the low excitation current (1 nA) used in the measurements. All data are acquired at $T=0.3$~K.
}
\label{fig:1}
\end{figure*}

Here, we report electrical transport measurements from three tMBG samples over a tight range of twist angles, $1.13^{\circ} \leq \theta \leq 1.19^{\circ}$. Notably, our studied range of twist angles falls intermediate to recent prior reports~\cite{Chen2020tMBG,Polshyn2020,Shi2020}, enabling a more complete understanding of the evolution of correlated states in tMBG with $\theta$. We primarily focus our attention on two devices with twist angles of $\theta=1.13^{\circ}$ (device D1) and $1.19^{\circ}$ (device D2). Figures~\ref{fig:1}a-b show maps of the longitudinal resistivity, $\rho_{xx}$, in device D1 for $D>0$ and $D<0$, respectively. The maps are primarily confined to the flat conduction band ($0 \leq \nu \leq 4$, where $\nu$ is the band filling factor as defined in Methods) and large $|D|$, for which correlated states are observed at low temperature. We assume a convention in which $D>0$ corresponds to the electric field pointing from the monolayer to the bilayer graphene. We observe robust insulating states at $\nu=0$ and $4$ for both signs of $D$, as anticipated from calculations of the single-particle band structure of tMBG. We additionally see well-developed correlated insulating states at $\nu=1$ and $2$ for $D>0$ (Fig.~\ref{fig:1}a), as well as correlated metallic states at $\nu=1$, $2$, and $3$ for $D<0$ (Fig.~\ref{fig:1}b). Here, we define insulating states as those exhibiting increasing $\rho_{xx}$ as the temperature is lowered, whereas metallic states behave oppositely (see Supplementary Information Figs.~\ref{fig:nu1_temperature}a-b and~\ref{fig:deviceD1}c-d for the temperature dependence of $\rho_{xx}$).

This correlated phase diagram connects smoothly to prior measurements of devices with slightly different twist angles. For $D>0$, devices over a wide range of twist angles manifest a robust correlated insulating state at $\nu=2$~\cite{Chen2020tMBG,Polshyn2020,Shi2020}. However, the states at $\nu=1$ and $3$ are absent in devices with slightly smaller twist angles (although appear to re-emerge in a device with even smaller twist angle, $\theta=0.89^{\circ}$~\cite{Chen2020tMBG}, subsequent to an anticipated topological transition in the band from $C_v=2$ to $1$~\cite{Park2020theory}). In contrast, both of these states are seen in devices with slightly larger twist angles~\cite{Polshyn2020,Shi2020}, before eventually disappearing again in devices with $\theta \gtrsim 1.4^{\circ}$~\cite{Chen2020tMBG,Polshyn2020,Shi2020}. The correlated states at $D<0$ are less sensitive to twist angle, with resistive bumps observed at all integer $\nu$ in devices with $1.05^{\circ} \lesssim \theta \lesssim 1.4^{\circ}$~\cite{Chen2020tMBG,Polshyn2020,Shi2020}. The absence of robust correlated insulating states is likely a consequence of the larger bandwidth compared to the $D>0$ bands~\cite{Ma2020,Park2020theory,Rademaker2020}. Supplementary Information Section S4 and Table~\ref{tab:summary} provide a detailed summary of the correlated states observed in our three devices, as well as those previously reported in Refs.~\cite{Chen2020tMBG,Polshyn2020,Shi2020}.

\begin{figure*}[t]
\includegraphics[width=5.5 in]{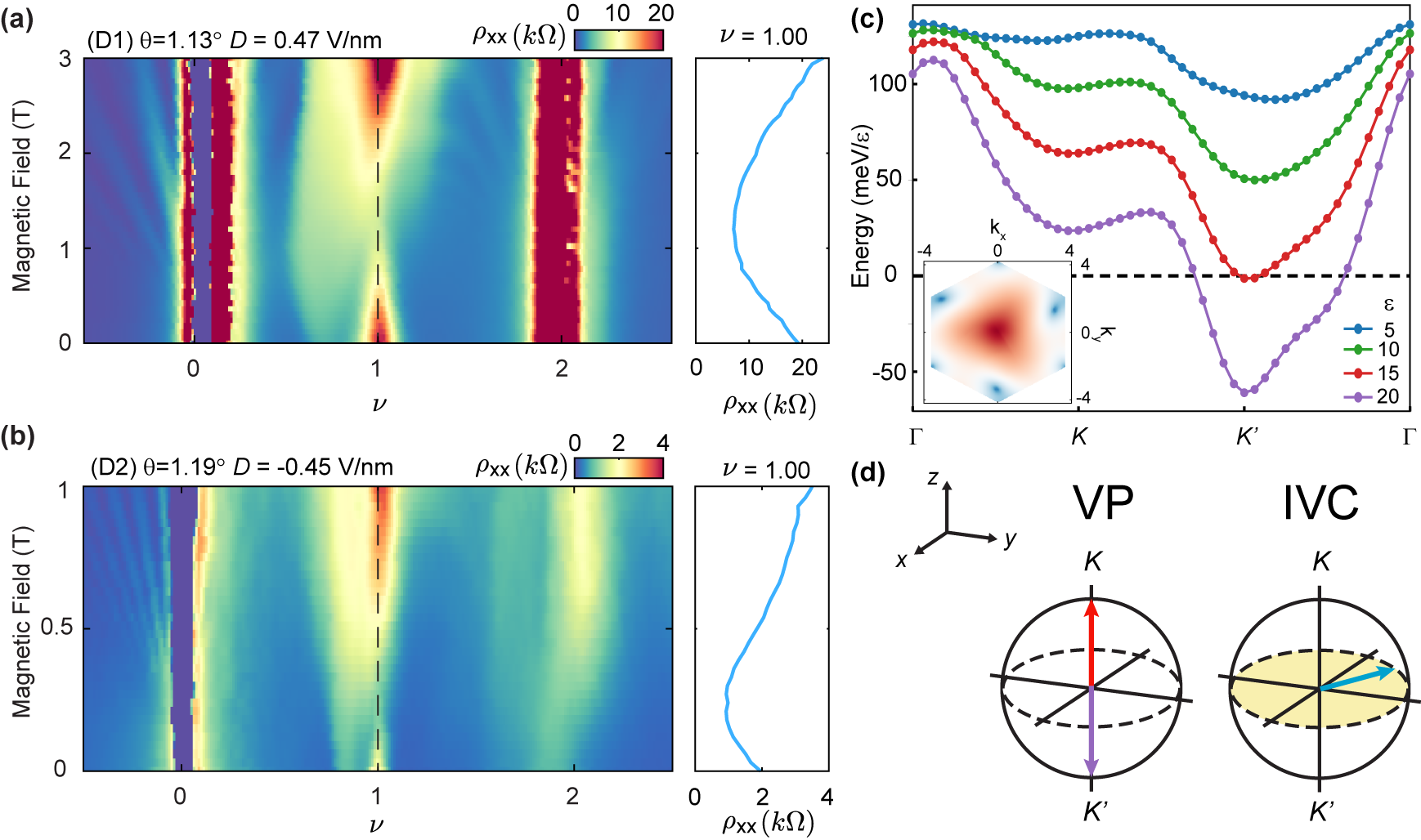} 
\caption{\textbf{Magnetic field dependence and ground state ordering of the $\nu=1$ state.}
\textbf{a-b}, $\rho_{xx}$ as a function of doping and $B$ at $T=0.1$~K for $D>0$ in device D1 (\textbf{a}) and at $T=0.05$~K for $D<0$ in device D2 (\textbf{b}). Cuts of $\rho_{xx}(B)$ are shown at $\nu=1$ in the panels to the right of each map, at positions indicated by the black dashed lines in the main panels. Note that $\nu$ is the fast sweeping axis in these measurements, and as a consequence the measured $\rho_{xx}$ is substantially smaller than its true value for the insulating states; Supplementary Information Fig.~\ref{fig:nu1_absentAHE}a shows a more faithful measurement of $\rho_{xx}$ at $\nu=1$ in device D1.
\textbf{c}, Calculated energy of valley magnon formation in tMBG with $\theta=1.16^{\circ}$ at selected values of $\epsilon$. The inset shows the profile of the intervalley exciton, $F(\mathbf{k})$, as a function of crystal momentum in the first Brillouin zone, with the magnitude represented by a log color scale. $k_{x,y}$ have units $1/a_M$, where $a_M$ is the moir\'e lattice constant.
\textbf{d}, Bloch sphere representation of the VP and IVC states. VP states point towards the north (south) pole for K (K') polarization, as indicated by the red (purple) arrows. IVC states (blue arrow) point along any direction in the $x$-$y$ plane (shaded).
}
\label{fig:2}
\end{figure*}

In all three of our devices, we observe an AHE within the ``halo'' region associated with the symmetry-broken state at $\nu=1$ for $D>0$ (see Figs.~\ref{fig:1}c-d and Supplementary Information Figs.~\ref{fig:deviceD2_AHE} and~\ref{fig:deviceD3}). We see hysteretic behavior in both $\rho_{xx}$ and $\rho_{yx}$ as $B$ is swept back and forth at fixed $\nu$ and $D$ (Figs.~\ref{fig:1}c-d for device D1). Because spin-orbit coupling is extremely weak in graphene, spin-ordered magnetism is not anticipated to result in an AHE in tMBG. Instead, the AHE with hysteretic behavior is very likely tied to orbital magnetism~\cite{Sharpe2019,Serlin2020,Chen2020,Tschirhart2020,Chen2020tMBG,Polshyn2020}. Figures~\ref{fig:1}e-f show maps of the AHE versus $\nu$ at two different fixed $D>0$, acquired by taking the difference of $\rho_{yx}$ between the two field sweeping directions, $\Delta \rho_{xy} = \rho_{yx}^{B\uparrow}-\rho_{yx}^{B\downarrow}$. Precisely at $\nu=1$ in the map acquired at $D=0.47$~V/nm (Fig.~\ref{fig:1}e), we measure a large $\rho_{xx}$ of tens of kiloohms and a rapidly oscillating $\Delta \rho_{yx}$ consistent with the behavior of a trivial insulating state, rather than a quantum anomalous Hall (QAH) state (see also Supplementary Information Figs.~\ref{fig:nu1_absentAHE}a-b). This contrasts prior reports in devices with $\theta=1.25^{\circ}$, in which a nearly quantized AHE in $\rho_{yx}$ and a small $\rho_{xx}$ are observed precisely at $\nu=1$~\cite{Polshyn2020}. For $D<0$, our measurements additionally reveal large pockets of AHE near $\nu=1$ and $3$ (Figs.~\ref{fig:1}g-j), which have previously not been reported.

\begin{figure*}[t]
\includegraphics[width=6.9 in]{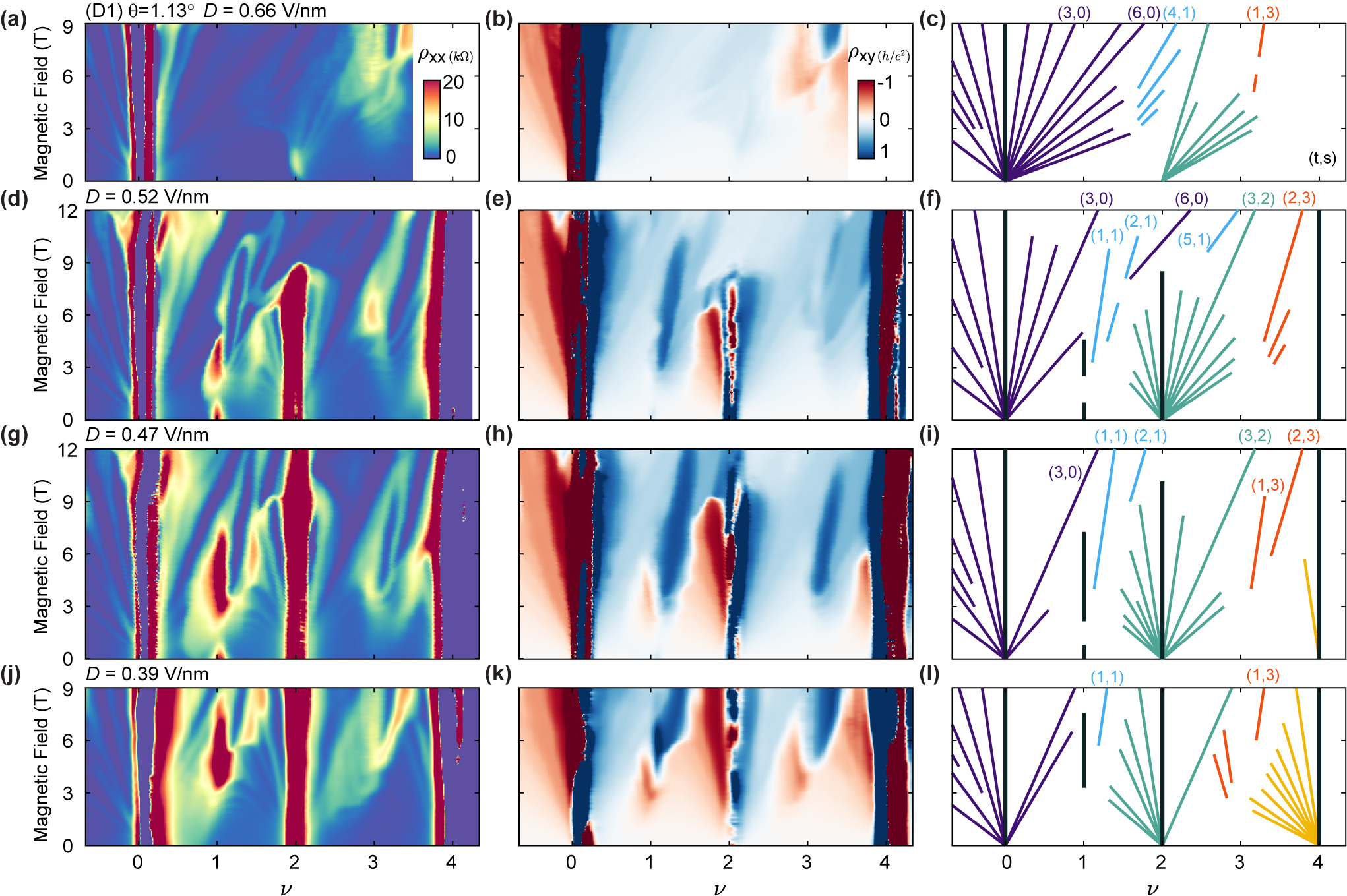} 
\caption{\textbf{Landau fan diagrams and spontaneous flavor polarization at high field for $D>0$.}
Landau fan diagrams in device D1 at \textbf{a-c}, $D=0.66$~V/nm, \textbf{d-f}, $D=0.52$~V/nm, \textbf{g-i}, $D=0.47$~V/nm, and \textbf{j-l}, $D=0.39$~V/nm, all acquired below $T=0.3$~K. The leftmost column shows $\rho_{xx}$, the central column shows $\rho_{xy}$, and the rightmost column schematically denotes the strongest observed gapped states. In the schematic, purple lines correspond to states tracing to $\nu=0$, blue to $\nu=1$, green to $\nu=2$, orange to $\nu=3$, and yellow to $\nu=4$. The vertical black lines denote topologically trivial insulating states. Selected states are labeled by their respective $(t,s)$ indices.
}
\label{fig:3}
\end{figure*}

The AHE has been observed previously in a number of graphene-based moir\'e platforms (tBLG~\cite{Sharpe2019,Serlin2020}, aligned ABC trilayer graphene on boron nitride~\cite{Chen2020}, and tMBG~\cite{Chen2020tMBG,Polshyn2020}), with a tendency towards quantization precisely at odd integer $\nu$ ($1$ and/or $3$). In the simplest case, a QAH state in tMBG is expected to arise upon polarization into a single spin- and valley-polarized band precisely at odd integer $\nu$ owing to the finite $C_v=2$ of the conduction band. Quantization is lost upon doping with electrons or holes, however an AHE can still persist owing to the large Berry curvature at the band extrema. Upon applying a magnetic field, the trajectory of the state drifts in the $n-B$ phase space as described by the St\v{r}eda formula~\cite{Streda1982}, $C=(h/e)(\partial n/\partial B)$, with slope equal to the Chern number, $C$, of the state ($h$ is Planck's constant, $e$ is the charge of the electron, and $n$ is the charge carrier density). All of these features have been observed previously in tMBG devices for $D>0$ at slightly larger twist angle ($\theta=1.25^{\circ}$) with $C=2$~\cite{Polshyn2020}. Our results in devices with slightly smaller twist angles contrast these expectations, however, exhibiting neither signatures of a QAH state precisely at odd integer $\nu$, nor the anticipated finite sloping of these states in a weak magnetic field. Figure~\ref{fig:2}a shows the low-field Landau fan diagram at $D>0$ in device D1, along with a cut of $\rho_{xx}$ versus $B$ taken precisely at $\nu=1$ in the panel to the right. We find that the insulating state at $\nu=1$ projects vertically and is suppressed with small $B$, before eventually re-emerging at larger $B$. Although in principle a disordered network of VP states with nearly equal mixture of valley K and K' domains may localize to form a trivial insulator, the application of a weak $B$ should rapidly align the domains and form a Chern insulator state. Our observations are inconsistent with this scenario, suggesting that the $B=0$ ground state is not a VP state.

\begin{figure*}[t]
\includegraphics[width=6.9 in]{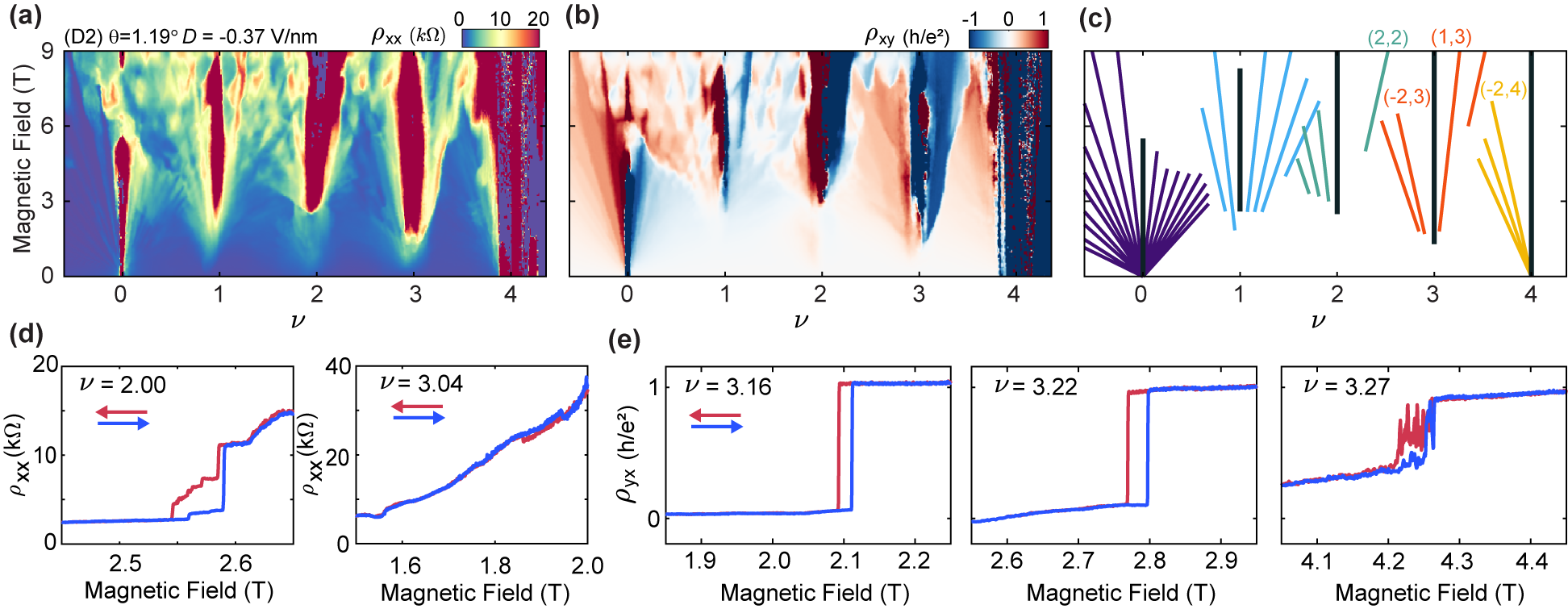} 
\caption{\textbf{First-order orbital phase transitions at high field for $D<0$.}
\textbf{a-b}, Landau fan diagram of $\rho_{xx}$ (\textbf{a}) and $\rho_{xy}$ (\textbf{b}) for device D2 at $D=-0.37$~V/nm, acquired at $T=0.1$~K.
\textbf{c}, Schematic representation of the observed states, following the convention established in Fig.~\ref{fig:3}. A subset of the most robust gapped states we observe are labeled by their respective $(t,s)$ indices. States at all integer $t$ are observed for $s=0$ and $1$, indicating full degeneracy lifting.
\textbf{d}, $\rho_{xx}$ acquired as $B$ is swept back and forth at $\nu=2.00$ (left) and $\nu=3.04$ (right).
\textbf{e}, $\rho_{yx}$ acquired as $B$ is swept back and forth at $\nu=3.16$ (left), $\nu=3.22$ (middle), and $\nu=3.27$ (right).
Data in \textbf{d-e} are acquired at $T=0.1$~K.
}
\label{fig:4}
\end{figure*}

Figure~\ref{fig:2}b shows a similar measurement for $D<0$ in device D2. This device is notable in that it exhibits the only known instance of insulating temperature dependence in a $D<0$ correlated state, in this case arising at $\nu=1$ (see temperature dependence in Supplementary Information Fig.~\ref{fig:nu1_temperature}c). Similar to the case of the $\nu=1$ state in device D1 at $D>0$, we find that $\rho_{xx}$ projects vertically, first becoming less resistive with $B$ before eventually growing at high field. Both of these insulating states at $B=0$ are accompanied by an AHE in a small region of $\nu<1$, approximately corresponding to the regions of sharply enhanced resistivity in Figs.~\ref{fig:2}a-b. This is consistent with our understanding that the AHE is associated with the symmetry-broken state at $\nu=1$, which persists well away from integer filling. Similar to the case of the $\nu=1$ state for $D>0$, we find that the AHE is strongest for $\nu \sim 0.9$ but vanishes at $\nu=1$, where a trivial insulating state emerges instead (see Supplementary Information Fig.~\ref{fig:deviceD2_AHE}e-f). Similarly, this state is not consistent with full valley polarization.

A number of correlated ground states have previously been proposed for twisted double bilayer graphene (tDBG)~\cite{Lee2019} (see Methods), and are candidate ground states for tMBG as well. Among those that result in an insulating state at odd integer $\nu$, none naturally explain our observation of an AHE upon doping, as this requires a highly unusual combination of broken time reversal symmetry (TRS), finite Berry curvature, but a net $C=0$. In order to gain more insight, we start with a fully spin- and valley-polarized ground state at $\nu=1$ and determine its stability by calculating the dispersion of the collective spin and valley wave excitation (see Methods and Supplementary Information Sections S1-S3 for full details). Negative energy of this excitation is a signature of the instability of the spin-valley polarized state. Figure~\ref{fig:2}c shows the calculated valley magnon energy for various values of the dielectric constant, $\epsilon$, which acts as an effective tuning parameter for the ratio $U/W$, where $U$ is the Coulomb interaction strength. We define a valley magnon as a spin-singlet valley-flip exciton (i.e. an exciton comprising two particles from opposite valleys). Condensation into an IVC order is favored when the valley magnon energy becomes negative at any point within the first Brillouin zone. We find that the VP state is stable at small $\epsilon$ (i.e. large $U/W$), however there is a first-order phase transition into an IVC order above a modest critical $\epsilon$ (i.e. intermediate $U/W$). Figure ~\ref{fig:2}d shows schematic representations of the VP and IVC states on the Bloch sphere.

Generically, the IVC state will retain TRS since it is a superposition of states at valleys K and K', trivializing the overall Chern number. However, our analysis of the IVC order parameter reveals that it prefers to carry finite angular momentum owing to the non-zero $C_v$ of the constituent bands (Fig.~\ref{fig:2}c inset). In combination with the large anticipated band edge Berry curvature, this IVC order is expected to result in a trivial insulating state at integer $\nu$ when gapped, but an AHE upon weakly doping the band into a metallic state. Although a full theoretical analysis is beyond the scope of this work, we find the high-angular momentum IVC state to be a plausible ground state order that is consistent with all of our observations near $\nu=1$ for both signs of $D$ (see Methods for a discussion of alternative plausible ground states).

Determining the ground state order for the correlated metallic states at $D<0$ is more challenging, since transport measurements cannot directly probe the topology of the state in absence of a gap. Since we do not observe any insulating behavior for the $D<0$ correlated states in device D1, the AHE we observe at $\nu=1$ and $3$ are consistent with either an ungapped VP state or an ungapped IVC state with finite angular momentum. The combination of larger $W$ and a smaller predicted $C_v=1$ naively favors IVC ordering even more strongly for the $D<0$ band compared with the $D>0$ band, which is thought to have $C_v=2$ (Supplementary Information Sections S2-S3). However, future work will be necessary to unambiguously identify these ground state orders.

The application of a large magnetic field serves to further modify the competition between different correlated ground states, both by tuning the orbital and spin Zeeman energies, and by transforming the low energy bands into a recursive series of Hofstadter minibands. Figure~\ref{fig:3} shows Landau fan diagrams in device D1 for both $\rho_{xx}$ and $\rho_{xy}$ at different values of $D>0$ (leftmost and middle columns, respectively). The schematics (rightmost column) denote the well-developed gapped states observed in each map. All are anticipated within a Hofstadter butterfly picture, in which gapped states follow trajectories described by the Diophantine equation, $\nu=tn_{\phi}+s$, where $t,s \in \mathbb{Z}$, and $n_{\phi}=\Phi/\Phi_{0}$ is the normalized magnetic flux. We refer to gapped states using the notation $(t,s)$, where $t$ corresponds to the Chern number of the state and $s$ corresponds to the number of electrons bound to each moir\'e unit cell. By convention, $(|t|>0,s=0)$ states are referred to as ``integer quantum Hall states,'' whereas $(|t|>0,s>0)$ are ``Chern insulators''~\cite{Spanton2018}. States with different $s$ are distinguished by color in the schematics, whereas trivial insulating states ($t=0$) are depicted in black irrespective of their corresponding $s$. 

We observe numerous similarities between the gapped states at filling factors $0 \leq \nu < 2$ and $2 \leq \nu \leq 4$. Although far from exact, there is an approximate mapping of our observed states between $\nu$ and $\nu+2$, especially for maps in which we observe a robust correlated insulating state at $\nu=2$ (i.e. Figs.~\ref{fig:3}d-l). The state at $\nu=2$ is thought to be spin-polarized at $B=0$~\cite{Chen2020tMBG}, suggesting that interactions split the four-fold degenerate conduction band into two sets of spin-polarized but valley-unpolarized bands over a wide range of $D>0$. The remaining valley degeneracy can also be spontaneously broken, however this often requires the assistance of a finite $B$ depending on $\nu$ and $D$ for our studied range of twist angles. Symmetry-broken states only persist to $B=0$ around $\nu=1$ for a small range of $D$, and we do not observe $B=0$ symmetry breaking at $\nu=3$ within our studied range of $\theta$. However, we find that numerous Chern insulator states emerge spontaneously at finite $B$ over a wide range of $D$ with both $s=1$ and $3$. These states are expected to be especially strong given the finite $C_v=2$ of the band, consistent with our observations and very likely indicative of a full flavor polarization of the $\nu=1$ and $3$ states at high field. In this context, the reentrant insulating behavior observed over a small range of $D$ at $\nu=1$ (Figs.~\ref{fig:3}d,g), along with the strong $(t>0,1)$ Chern insulators at high field, is consistent with a phase transition from an IVC state at low field to a VP state at high field.

More generally, we find that the states we observe in Fig.~\ref{fig:3} do not follow a simple progression with $\nu$ and $B$, reflecting the rich competition between the gapped single-particle Hofstadter subbands and the tendency towards spontaneous flavor polarization into a subset of these bands (see Methods for additional discussion). This is similar to recent observations in tBLG~\cite{Nuckolls2020,Wu2020,Saito2020,Das2020,Choi2020,Park2020}, however here we observe a more complicated sequence of symmetry breaking in which various flavor-polarized and unpolarized ground states closely compete. For example, in Figs.~\ref{fig:3}d-f, the main-sequence quantum Hall state emanating from the charge neutrality point, $(6,0)$, clearly intercedes both the $(2,1)$ and $(0,2)$ states and closes those gaps at high field. In particular, the $(2,1)$ gap closes and then reopens at higher field following this interruption. Additionally, a $(5,1)$ state is observed at filling factors $\nu>2$ for $B \gtrsim 9$~T. Tuning $D$ further tips the balance in the competition between these states, highlighting their near degeneracy.

For $D<0$, we observe different manifestations of field-driven competitions between correlated ground states. Figures~\ref{fig:4}a-b show Landau fan diagrams for $\rho_{xx}$ and $\rho_{xy}$ at $D=-0.37$~V/nm in device D2, with an associated schematic of the observed gapped states shown in Fig.~\ref{fig:4}c. Consistent with prior observations~\cite{Chen2020tMBG}, we see the emergence of correlated insulating states at all integer $\nu$ at finite $B$. However, the improved quality of this device reveals a number of previously obscured features of these states. We observe abrupt transitions from metallic to insulating states at $\nu=2$ and $3$ above a critical magnetic field, $B_{c}$, as shown in Fig.~\ref{fig:4}d. We additionally observe hysteresis at high field upon sweeping $B$ back and forth across this phase transition at $\nu=2$ (as well as very weak hysteresis signatures at $\nu=3.04$), indicative of a first-order phase transition. The hysteresis additionally extends to band fillings well away from integer $\nu$. For example, Fig.~\ref{fig:4}e shows $\rho_{yx}$ at various $\nu>3$, in which we see a hysteretic transition to a state that is nearly quantized to $h/e^2$ above $B_{c}$. Additional measurements in device D1 show a $(1,3)$ Chern insulator that persists nearly to $B=0$ (see Supplementary Information Figs.~\ref{fig:deviceD1}e-g), suggesting that the high-field state is likely fully flavor polarized owing to the anticipated $C_v=1$ of the band. The hysteresis we observe in Fig.~\ref{fig:4}e persists to values of $\nu$ for which the gapped states at $B<B_{c}$ carry $s=4$ (see Fig.~\ref{fig:4}c), indicating a first-order topological phase transition between the flavor-unpolarized Fermi surface near full band filling and the (presumably) fully flavor-polarized Fermi surface associated with $\nu=3$.

Overall, our results reveal the richness of the correlated phase diagram of tMBG, which can be tuned sensitively with the combination of $\nu$, $D$, $B$, and $\theta$. Our observation of a potential IVC state at odd integer $\nu$ appears to cede to a VP state for $D>0$ in devices with slightly larger twist angles~\cite{Polshyn2020}, whereas flavor-unpolarized states are observed in devices with slightly smaller twist angles~\cite{Chen2020tMBG}. This suggests that the ground state order may be controlled sensitively by $U/W$, in which sequential transitions from VP to IVC to flavor-unpolarized states are driven by an increasing $W$ with reducing $\theta$. Magnetic field further tunes the close competition between flavor-unpolarized ground states and numerous flavor-polarized states, resulting in abrupt and occasionally hysteretic topological phase transitions at high field.  

\section*{Methods}

\textbf{Device fabrication.} tMBG devices were fabricated using the ``cut-and-stack'' method~\cite{Chen2019a,Saito2019}, in which exfoliated graphene flakes with connected monolayer and bilayer regions are isolated using an atomic force microscope tip, and then stacked atop one another at the desired twist angle. Samples were assembled using standard dry-transfer techniques with a polycarbonate (PC)/polydimethyl siloxane (PDMS) stamp~\cite{Wang2013}. All tMBG devices are encapsulated in flakes of BN and graphite, and then transferred onto a Si/SiO$_2$ wafer. The temperature was kept below 180$^{\circ}$C during device fabrication to preserve the intended twist angle. Standard electron beam lithography, CHF$_3$/O$_2$ plasma etching, and metal deposition techniques (Cr/Au) were used to define the complete stack into Hall bar geometry~\cite{Wang2013}.

\textbf{Transport measurements.} Transport measurements were performed in a Bluefors dilution refrigerator with heavy low temperature electronic filtering, and were conducted in a four-terminal geometry with a.c. current excitation of 1-10 nA using standard lock-in techniques at a frequency of 13.3 Hz. In some cases, a gate voltage was applied to the Si gate in order to dope the region of the graphene contacts overhanging the graphite back gate to a high charge carrier density and reduce the contact resistance. $n$ and $D$ could be tuned independently with a combination of the top and bottom graphite gate voltages through the relations $n=(V_{t}C_{t}+V_{b}C_{b})/e$ and $D=(V_{t}C_{t}-V_{b}C_{b})/2\epsilon_0$, where $C_t$ and $C_b$ are the top and bottom gate capacitance, $V_t$ and $V_b$ are the top and bottom gate voltage, and $\epsilon_0$ is the vacuum permittivity.

\textbf{Twist angle determination.} The twist angle $\theta$ is first determined from the values of $n$ at which the insulating states at full band filling ($\nu=\pm4$) appear, following $n=8\theta^2/\sqrt{3}a^2$, where $a=0.246$~nm is the lattice constant of graphene. It is then confirmed by fitting the high field gapped states to a Wannier diagram anticipated from the Hofstadter butterfly spectrum. The filling factor is defined as $\nu=\sqrt{3}\lambda^2 n/2$, where $\lambda$ is the period of the moir\'e.

\textbf{Theoretical modeling of the $\nu=1$ ground state.} We calculate the band structure of tMBG using a standard continuum model, taking $\theta=1.16^{\circ}$ as a value intermediate to the twist angles of our measured devices. We project the Hamiltonian to include only the conduction band, and calculate the energy of a valley wave excitation assuming a fully spin-valley polarized ground state at $\nu=1$. We analyze the symmetry of the resulting IVC order depending on the value of $C_v$, and find that in general the profile of the intervalley exiton, $F(k)$, carries a total vorticity of $2C_v$, and consequentially is not constant in momentum space for $C_v>0$. This IVC state is degenerate with its time-reversed partner, forming a superposition that preserves the overall TRS. However, if this degeneracy is spontaneously lifted, the resulting ground state breaks TRS. This state is also relevant at $\nu=3$, however we do not observe a gap at that filling in any of the devices reported here, precluding a direct experimental comparison between the IVC and VP states. Detailed Hartree-Fock calculations can more faithfully assess the competition between the VP state and the various IVC orders, however they are complicated by uncertainties in band structure parameters and details of the relevant interactions in tMBG, and are beyond the scope of this work. Full details of our calculations can be found in Supplementary Information Sections S1-3. 

\textbf{Additional candidate ground states at $\nu=1$.} A more subtle origin of the behavior of the states at $\nu=1$ is that the ground state order changes upon doping. In this scenario, the ground state at and very near $\nu=1$ is an IVC that preserves TRS. Such an IVC state also trivializes the Chern number, resulting in a trivial insulating state when gapped. A phase transition to a fully flavor-polarized state arises upon doping away from $\nu=1$, leading to an AHE in the metallic states owing to the inherent TRS breaking and finite $C$. Theoretically, this phase transition is expected to be first order. Since we do not observe any obvious signatures of hysteresis with doping, our results are naively inconsistent with this scenario. However, it remains possible that this phase transition is smeared by various forms of disorder in the sample, including twist disorder. For this reason, we are not able to unambiguously distinguish between this scenario and the case of the TRS-breaking IVC order detailed in the main text. Although we believe it to be unlikely to achieve a first-order phase transition between an IVC and VP state with a small amount of doping, a more detailed theoretical analysis is necessary to assess its feasibility. We also discuss an additional form of density-wave IVC order consistent with our results in Supplementary Information Section S3. Finally, we note that we have not performed an exhaustive search for all possible ground states at $\nu=1$. Although there may be other states that are consistent with trivial insulating behavior at integer $\nu$ and an AHE upon doping, they are likely to be more exotic than the various IVC orders considered here.

\textbf{Determination of the valley Chern number.} Our calculated band structure has $C_v=2$ ($1$) for the $D>0$ ($D<0$) conduction band (Supplementary Information Fig.~\ref{fig:dispersion}). With full valley polarization at odd integer $\nu$, the gapped state at $B=0$ has $C=C_v$~\cite{Polshyn2020}. However, in the apparent absence of gapped VP states at $B=0$ in our devices, direct confirmation of the value of $C_v$ is no longer possible. Although we observe symmetry-broken states with apparent VP at large $B$, the magnetic field additionally transforms the conduction band into a recursive sequence of Hofstadter subbands that may carry different values of $C$ than at $B=0$. The strongest gapped states observed at high field are not necessarily constrained to a single value of $C$, but may in principle also depend dynamically on the doping since their trajectories differ according to the St\v{r}eda formula. For example, we observe comparably robust $(1,1)$ and $(2,1)$ states in Figs.~\ref{fig:3}d-f, and comparable $(1,3)$ and $(2,3)$ states in Figs.~\ref{fig:3}g-i. The abrupt emergence of these gapped states with $B$ provides strong evidence that they are driven by spontaneous flavor polarization, and the separation of states with different $t$ at high field permits multiple such correlated Chern insulators to coexist at a given $B$. Because we only observe these states at relatively high field for $D>0$, we are unable to directly verify the anticipated $C_v=2$ for the $B=0$ conduction band, although their existence implies that $C_v>0$. Generically, the $D<0$ states are similarly problematic, however in one instance we observe a robust $(1,3)$ state that emerges at relatively small $B$ in device D1 (Supplementary Information Fig.~\ref{fig:deviceD1}e-g). Although inconclusive without a corresponding QAH state at $B=0$, it is suggestive that the $D<0$ band has $C_v=1$ as anticipated.

\textbf{Relation to $\nu=1$ and $3$ states in tDBG.} The correlated phase diagram of tDBG exhibits many qualitative similarities with that of tMBG at $D>0$~\cite{Chen2020tMBG}, with a notable exception that $B=0$ Chern insulator states are observed in tMBG~\cite{Chen2020tMBG,Polshyn2020} but not in tDBG~\cite{Burg2019,Shen2020,Cao2020,Liu2020,He2020}. In particular, tDBG states at $\nu=1$ and $3$ are typically absent at $B=0$, or only weakly insulating at low temperatures in devices with $\theta \approx 1.23^{\circ} - 1.30^{\circ}$~\cite{Cao2020,He2020}. Supplementary Information Fig.~\ref{fig:tDBG_noAHE} shows transport measurements of a tDBG device with $\theta=1.30^{\circ}$, acquired at $T=50$~mK (note that this is the same as device D3 from Ref.~\cite{He2020}). We observe correlated insulating states and surrounding ``halo'' features at $\nu=1$ and $3$. The lowest moir\'e conduction band of tDBG is also expected to have $C_v=2$~\cite{Lee2019}, therefore the observation of trivial insulating states at $\nu=1$ and $3$ appears to be inconsistent with ground states that are both spin- and valley-polarized. Although this behavior is reminiscent of our observations for the $\nu=1$ state in our tMBG devices, we do not observe any signatures of the AHE in any of our measurements of $\rho_{xy}$ at or nearby these states in tDBG (Supplementary Information Figs.~\ref{fig:tDBG_noAHE}b-c). These observations could also be plausibly explained by some form of IVC order at $\nu=1$ and $3$. As discussed above, at least two different types of IVC states that retain TRS are possible: one with an $s$-wave intervalley exciton  profile, and another comprising a superposition of time-reversed pairs of IVC states with higher angular momentum. The preserved TRS of these states is necessary to explain the absence of the AHE upon doping. Additional work will be necessary to further interrogate the potential connections between these states and those we observe in tMBG.

\section*{acknowledgments}
We thank Shaowen Chen, Cory Dean, Andrea Young, Ashvin Vishwanath, and David Cobden for helpful discussions. Technical support for the dilution refrigerator was provided by A. Manna. This work was supported by NSF MRSEC 1719797 and the Army Research Office under Grant Number W911NF-20-1-0211. X.X. acknowledges support from the Boeing Distinguished Professorship in Physics. X.X. and M.Y. acknowledge support from the State of Washington funded Clean Energy Institute. This work made use of a dilution refrigerator system which was provided by NSF DMR-1725221. Y.H.L. acknowledges the support of the China Scholarship Council. K.W. and T.T. acknowledge support from the Elemental Strategy Initiative conducted by the MEXT, Japan, Grant Number JPMXP0112101001, JSPS KAKENHI Grant Number JP20H00354 and the CREST (JPMJCR15F3), JST.\\

\section*{Author contributions}
M.H. and Y.L. fabricated the devices. M.H. performed the measurements, with assistance from Y.L. and Z.F. Y.-H. Z. performed the theoretical calculations. K.W. and T.T. grew the BN crystals. M.H., X.X., and M.Y. analyzed the data and wrote the paper with input from all authors.

\section*{Data Availability}
Source data are available for this paper. All other data that support the plots within this paper and other findings of this study are available from the corresponding author upon reasonable request.

\section*{Competing interests}
The authors declare no competing interests.

\section*{Additional Information}
Correspondence and requests for materials should be addressed to X.X. or M.Y.

\section*{Supplementary Information}
Supplementary Sections S1-S4, Table S1, and Figs. S1-S9.

\bibliographystyle{naturemag}
\bibliography{references}

\clearpage


\renewcommand{\thefigure}{S\arabic{figure}}
\renewcommand{\thesubsection}{S\arabic{subsection}}
\setcounter{secnumdepth}{2}
\renewcommand{\theequation}{S\arabic{equation}}
\renewcommand{\thetable}{S\arabic{table}}
\setcounter{figure}{0} 
\setcounter{equation}{0}

\onecolumngrid

\section*{Supplementary Information}

\subsection{Band structure calculation and valley Chern number}

We calculate the band structure of tMBG using the standard continuum model. The Hamiltonian is
\begin{equation}
	H=H_{MG}+H_{BG}+H_M
\end{equation}

We have  
\begin{equation}
	H_{MG}=\sum_{\mathbf k}(\tilde c^\dagger_{A}(\mathbf k), \tilde c^\dagger_{B}(\mathbf k))\left(
	\begin{array}{cc}
	-\frac{\delta}{2}& -\frac{\sqrt{3}}{2}t(\tilde k_x-i \tilde k_y)\\
	-\frac{\sqrt{3}}{2}t(\tilde k_x+i \tilde k_y) &-\frac{\delta}{2} 
	\end{array}\right)  \left(\begin{array}{c} \tilde c_A(\mathbf k)\\ \tilde c_B(\mathbf k)\end{array}\right)
\end{equation}
where $\mathbf{\tilde k}=R(-\theta/2) \mathbf {k}$, with twist angle $\theta$. $R(\varphi)$ is the transformation matrix for anticlockwise rotation with angle $\varphi$.

The Hamiltonian for the bilayer graphene is
\begin{equation}
	H_{BG}=\sum_{\mathbf k}\Psi^\dagger(\mathbf k)\left(
	\begin{array}{cccc}
	0& -\frac{\sqrt{3}}{2}t(\tilde k_x-i\tilde k_y) & -\frac{\sqrt{3}}{2}\gamma_4 (\tilde k_x-i \tilde k_y) & -\frac{\sqrt{3}}{2}\gamma_3(\tilde k_x+i\tilde k_y)\\
	-\frac{\sqrt{3}}{2}t(\tilde k_x+i\tilde k_y) &0 & \gamma_1  & -\frac{\sqrt{3}}{2}\gamma_4 (\tilde k_x-i \tilde k_y)\\
	-\frac{\sqrt{3}}{2}\gamma_4(\tilde k_x+i\tilde k_y) & \gamma_1 & \frac{\delta}{2}& -\frac{\sqrt{3}}{2}t (\tilde k_x-i\tilde k_y)\\
	-\frac{\sqrt{3}}{2}\gamma_3(\tilde k_x-i\tilde k_y) &-\frac{\sqrt{3}}{2}\gamma_4(\tilde k_x+i\tilde k_y) & -\frac{\sqrt{3}}{2}t(\tilde k_x+i\tilde k_y) & \frac{\delta}{2}
	\end{array}\right)  \Psi(\mathbf k)
\end{equation}
where $\mathbf{\tilde k}=R(\theta/2)\mathbf{k}$ and $\Psi^\dagger(\mathbf k)=(c^\dagger_{A_1}(\mathbf k), c^\dagger_{B_1}(\mathbf k), c^\dagger_{A_2}(\mathbf k), c^\dagger_{B_2}(\mathbf k))$.

Finally the interlayer moir\'e tunneling term is
\begin{equation}
	H_M=\sum_{\mathbf k} \sum_{j=0,1,2} (\tilde c^\dagger_{A}(\mathbf k), \tilde c^\dagger_{B}(\mathbf k))\left(
	\begin{array}{cc}
	\alpha t_M& t_M e^{-i  \frac{2\pi}{3} j}\\
	t_M e^{i  \frac{2\pi}{3} j} &\alpha t_M 
	\end{array}\right)  \left(\begin{array}{c} c_{A_1}(\mathbf k+\mathbf{Q}_j)\\ c_{B_1}(\mathbf k+\mathbf{Q}_j)\end{array}\right)+h.c.
\end{equation}
where $\mathbf Q_0=(0,0)$, $\mathbf Q_1=\frac{1}{a_M}(-\frac{2\pi}{\sqrt{3}},-2\pi)$ and $\mathbf Q_2=\frac{1}{a_M}(\frac{2\pi}{\sqrt{3}},-2\pi)$, with $a_M$ as the moir\'e lattice constant.

We use parameters $(t, \gamma_1, \gamma_3, \gamma_4)=(-2610, 361, 283, 140)$ meV. For the interlayer tunneling, we use $t_M=110$ meV and $\alpha=0.5$. $\delta$ is the potential difference, which is tuned by the displacement field, $D$.
 
\begin{figure*}[h]
\includegraphics[width=5in]{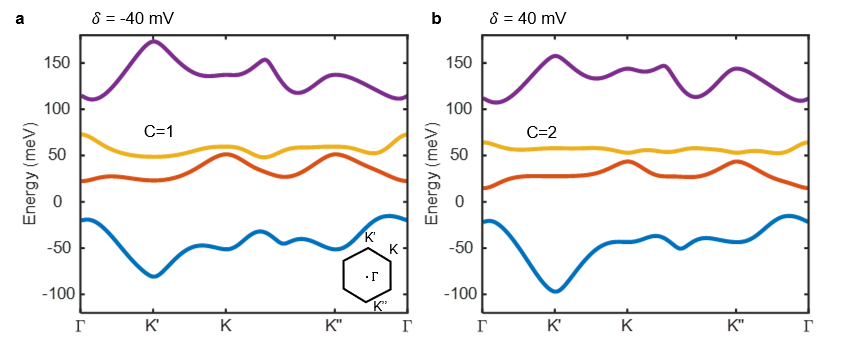}
\caption{\textbf{Band structure of tMBG at twist angle $\theta=1.16^\circ$.}
\textbf{a}, $\delta=-40$ meV and \textbf{b}, $\delta=40$ meV. The valley Chern number of the lowest moir\'e conduction band (yellow) is $C_v=1$ for $\delta=-40$~meV, and $C_v=2$ for $\delta=40$~meV. The inset in \textbf{a} shows the relevant high symmetry points in the first Brillouin zone.
}
\label{fig:dispersion}
\end{figure*}

The dispersion at $\delta=-40$ meV and $\delta=40$ meV for twist angle $\theta=1.16^\circ$ is shown in Fig.~\ref{fig:dispersion}. The valley Chern numbers, $C_v$, for the conduction band are $1$ and $2$, respectively. Note that we take a convention in which $\delta>0$ corresponds to the electric field pointing from the monolayer to the bilayer graphene (similarly, this corresponds to $D>0$).

\subsection{Calculation of valley wave dispersion}

For simplicity, we project the Hamiltonian to include only the conduction band. However, the other remote bands still play a role in renormalizing the dispersion of the active conduction band from the interaction:

\begin{align}
H_V= \frac{1}{2} \frac{1}{N_s} \sum_{\mathbf q}V(\mathbf q) \sum_{\mathbf k_1} \sum_{\mathbf k_2} c^\dagger_{a;m_1}(\mathbf k_1 +\mathbf q)c^\dagger_{b;n_1}(\mathbf k_2-\mathbf q) c_{b;n_2}(\mathbf k_2) c_{a;m_2}(\mathbf k_1) \lambda_{a;m_1 m_2}(\mathbf k_1, \mathbf q) \lambda_{b;n_1 n_2}(\mathbf k_2, -\mathbf q)
\end{align}
where,

\begin{equation}
	V(\mathbf q)=\frac{15 \text{nm}}{a_M} \frac{1392 \text{meV}}{\epsilon} \frac{1}{q a_M} \tanh(q r_0) 
\end{equation}
Here $r_0$ is the screening length from the metallic contact, which we use $r_0=5 a_M$. In the above $\lambda_a;{mn}(\mathbf k, \mathbf q)= \braket{ \mu_{a;m}(\mathbf k+\mathbf q)|\mu_{b;n}(\mathbf k)}$ is the form factor. Here $a,b=+,-$ is the valley index and $m,n$ is the band index.

Let us fully fill all of the bands below the conduction band. Then the kinetic term of the conduction band is renormalized:
\begin{equation}
	H_K=H^0_K+ \sum_{\mathbf k}\xi_{H}(\mathbf k) c^\dagger(\mathbf k) c(\mathbf k)+ \sum_{\mathbf k}\xi_F (\mathbf k) c^\dagger(\mathbf k) c(\mathbf k)
\end{equation}
where $c$ is the operator for the valley $+$ of the conduction band.  $\epsilon_H$ is from the Hartree term and $\epsilon_F$ is from the Fock term.

We have
\begin{equation}
	\xi_{H}(\mathbf k)= \sum_{\mathbf G_M} V(\mathbf G_M)\rho(\mathbf G_M) \lambda_+(\mathbf k, \mathbf G_M)
\end{equation}
where,
\begin{equation}
	\rho(\mathbf G_M)=4 \frac{1}{N_s} \sum_{m\in O}  \lambda_{a;mm}(\mathbf k, \mathbf G_M)
\end{equation}
Here the factor $4$ comes from spin and valley degeneracy. $O$ is the set of occupied bands.

The Fock term is
\begin{equation}
	\xi_F(\mathbf k)=- \frac{1}{N_s}\sum_{\mathbf q} V(\mathbf q) \sum_{m \in O}   |\lambda_{a;mc}(\mathbf k, \mathbf q)|^2 
\end{equation}
where $m \in O$ is the index of the occupied band. $c$ is the index of the conduction band.   

Next, we calculate the valley wave assuming that the ground state is spin-valley polarized at $\nu=1$. Without loss of generality, we assume the ground state fully occupies $+,\downarrow$. A collective excitation with momentum $\mathbf{q}$ is generated by a particle-hole boson $b_{ab;\mathbf{q}}(\mathbf k)^\dagger=c^\dagger_{a;\uparrow}(\mathbf {k+q}) c_{b;\downarrow}(\mathbf k)$, where $a,b=+,-$ labels the valley.  Due to the $SU(2) \times SU(2)$ symmetry, the dispersion of the exciton corresponding to $c^\dagger_{+;\sigma}(\mathbf k+\mathbf q) c_{-;\sigma'}(\mathbf k)$  does not depend on the spin index $\sigma \sigma'$. In the following we will only consider the spin-singlet valley-flip exciton.

For a fixed momentum $\mathbf{q}$, we can derive Hamiltonian for $b_{ab;\mathbf{q}}(\mathbf k)$:
\begin{equation}
	H_{ab}(\mathbf q)=H^V_{ab}(\mathbf q)+H^K_{ab}(\mathbf q)
\end{equation}
where,
\begin{align}
	H^V_{ab}(\mathbf{q})&=- \sum_{\mathbf k \in MBZ} \sum_{\mathbf{\tilde q}}V(\mathbf{\tilde q})\big(\lambda_a(\mathbf{k+ q},\mathbf{\tilde q})\lambda_b(\mathbf{k+\tilde q},-\mathbf{\tilde q})  b^\dagger_{ab;\mathbf{q}}(\mathbf{k+\tilde q})b_{ab;\mathbf{q}}(\mathbf{k}) \notag\\
	&+\sum_{\mathbf{\tilde q}} V(\mathbf{\tilde q})\lambda_b(\mathbf{k},\mathbf{\tilde q})\lambda_b(\mathbf{k+ \tilde q},-\mathbf{\tilde q})  b^\dagger_{ab;\mathbf{q}}(\mathbf{k})b_{ab;\mathbf{q}}(\mathbf{k})  \notag\\
\end{align}
and
\begin{equation}
	H^K_{ab}(\mathbf q)=\sum_{\mathbf k \in MBZ} (\xi_{a}(\mathbf{k+q})-\xi_{b}(\mathbf k))b^\dagger_{ab;\mathbf{q}}(\mathbf{k})b_{ab;\mathbf{q}}(\mathbf{k})
\end{equation}
where $\xi_a$ is the dispersion after including the renormalization from remote bands.

For a fixed momentum $\mathbf{q}$, the ground state of $H_{ab}(\mathbf q)$ gives the energy of the excitation, $\omega_{ab}(\mathbf q)$. Here $\omega_{++}(\mathbf q)$ is the spin wave excitation and $\omega_{+-}(\mathbf q)$ is the spin-valley wave excitation.

We focus on the intervalley exciton $\omega_{+-}(\mathbf q)$. As shown in Fig.~\ref{fig:2}c of the main text, when we increase the dielectric constant $\epsilon$, the minimum energy of the exciton becomes negative at momentum $\mathbf q=K'$ in the mini Brillouin Zone (MBZ), suggesting an instability of the polarized ground state towards an intervalley coherent (IVC) state.

\subsection{Symmetry analysis of IVC order}
As discussed above, the valley polarized ground state is unstable because the energy of the valley magnon (or exciton) becomes negative when we increase the dielectric constant $\epsilon$. The condensation of valley magnons will lead to an IVC state. Here we perform a general symmetry analysis of the possible IVC states.

We consider a general IVC ansatz described by a mean field theory:
\begin{equation}
	H_M= - \sum_{\mathbf k}  [F(\mathbf k)c^\dagger_{+}(\mathbf k+\frac{1}{2}\mathbf Q) c_{-}(\mathbf k-\frac{1}{2}\mathbf{Q})+F^*(\mathbf k)c^\dagger_{-}(\mathbf k-\frac{1}{2}\mathbf Q) c_{+}(\mathbf k+\frac{1}{2}\mathbf Q)]
\end{equation}
where $F(\mathbf k)$ represents the profile of the intervalley exciton, similar to $\Delta(\mathbf k)$ in Cooper pairing. Here we assume that the exciton is condensed at momentum $\mathbf Q$, which may not be zero.

Using $T c_{+}(\mathbf k) T^{-1}=c_{-}(-\mathbf k)$ and $T c_{-}(\mathbf k) T^{-1}=c_{+}(-\mathbf k)$, we find:
\begin{equation}
	T H_M T^{-1}=-\sum_{\mathbf k}[F^*(\mathbf k) c^\dagger_{-}(-\mathbf k-\frac{1}{2}\mathbf Q) c_{+}(-\mathbf k+\frac{1}{2}\mathbf Q)+F(\mathbf k) c^\dagger_{+}(-\mathbf k+\frac{1}{2}\mathbf Q) c_{-}(-\mathbf k-\frac{1}{2}\mathbf Q)
\end{equation}

Then the time reversal symmetry $H_M=T H_M T^{-1}$ requires
\begin{equation}
	F(\mathbf k)= F(-\mathbf k)
\end{equation}

For an IVC state in the case of non-zero $C_v$, $F(\mathbf k)$ needs to acquire the Berry phase of the two bands with opposite signs of the Chern number. A general argument shows that the complex field $F(\mathbf k)$ cannot be a constant in momentum space, and should have total vorticity equal to $2C_v$. As a result, one usually finds that $F(\mathbf k)\neq F(-\mathbf k)$. This is reflected in our direct calculation of the intervalley exciton wavefunction on top of the spin-valley polarized ground state (see the inset of Fig.~\ref{fig:2}c of the main text). As a consequence, the IVC order breaks the time reversal symmetry. 

Let us also comment on the translation symmetry if $\mathbf Q \neq 0$. If only the IVC corresponding to one momentum $\mathbf Q$ is condensed, the resulting phase still preserves the moir\'e translation symmetry because we can apply a global phase generated by the valley charge. We call the IVC in this class a uniform IVC. However, if the IVC at both $\mathbf Q$ and $-\mathbf Q$ condense, the final state breaks the moir\'e translation symmetry and there should be a charge density wave (CDW) order at momentum $2\mathbf Q$. We call the IVC in this class a density-wave IVC. Note that the time reversal symmetry does not change $\mathbf Q$, thus either the uniform IVC or density-wave IVC breaks the time reversal symmetry if $F(\mathbf k)\neq F(-\mathbf k)$. A careful Hartree-Fock analysis is necessary to determine whether the uniform IVC or density-wave IVC is favored, which we leave to future work.  

\subsection{Summary of observed correlated states in tMBG}

Table~\ref{tab:summary} summarizes the correlated states observed at all integer $\nu$ in the conduction band of tMBG for both signs of $D$ in devices over a range of twist angles, $0.89^{\circ} \leq \theta \leq 1.385^{\circ}$. The data are compiled from devices D1, D2, and D3 reported here, as well as from devices D1 and D2 of Ref.~\cite{Chen2020tMBG} and devices D1, D3, and D4 of Ref.~\cite{Polshyn2020}. The rotational alignment of the tMBG to the top and bottom encapsulating BN is a latent parameter. We have no reliable estimate of these twist angles for any of our devices, nor do we see obvious signatures of close rotational alignment, however this may nevertheless play a role in driving variations between tMBG devices with otherwise comparable twist angles. 

We now comment on a number of details of the collectively reported tMBG devices. First, we note that in principle it possible that devices with $\theta \geq 1.25^{\circ}$ from Refs.~\cite{Polshyn2020} and~\cite{Shi2020} also exhibit weak AHE around $\nu=1$ and $3$ for $D<0$ states, however such measurements are not reported in those manuscripts, so we are unable to make such a determination. We also note that a second device with $\theta=1.25^{\circ}$ was also reported in Ref.~\cite{Polshyn2020}, exhibiting a very similar phase diagram to the first. However, measurements of the AHE are not shown, and we therefore omit it from the table. Finally, we note that devices with larger twist angles up to $\sim1.6^{\circ}$ have exhibited signatures of symmetry broken states for $D>0$~\cite{Chen2020tMBG,Shi2020}, however these states all exhibit metallic temperature dependence and have not been investigated in detail. We also omit these devices from the table.

Although identical sets of measurements have not been performed for all devices, the table appears to reveal a number of trends in the evolution of the correlated states with $\theta$. For $D>0$, the $\nu=2$ state is a robust trivial insulator over a wide range of twist angles, and appears to be spin-polarized from measurements of the gap with in-plane magnetic field, $B_{||}$~\cite{Chen2020tMBG}. A Chern insulator state at $B=0$ appears to emerge around $\nu=3$ only for a small range of twist angles near $1.25^{\circ}$. The AHE appears to approach a quantization at $h/2e^2$, suggesting the band has $C_v=2$ at this twist angle. For other twist angles, symmetry-broken states at $\nu=3$ are only observed at finite $B$. The $\nu=1$ state exhibits more complicated behavior. A gapped state is observed in one device with $\theta=0.89^{\circ}$~\cite{Chen2020tMBG}, but not in another at $0.90^{\circ}$~\cite{Polshyn2020}. In the former, a large AHE is observed, with quantization that appears to approach $h/e^2$, suggesting the band has $C_v=1$ at this twist angle. This behavior is consistent with a theoretically anticipated topological transition with twist angle around $\theta=1.0^{\circ}$~\cite{Park2020theory}. The gap closes as the twist angle increases to around $1.05^{\circ}$, before reemerging again in devices with twist angles larger than $1.13^{\circ}$. The unusual behavior surrounding this state has been discussed at length in the main text, but in short appears to be consistent with a transition from an IVC to VP state with increasing twist angle. The absence of a gapped state at slightly smaller twist angles of $\sim 1.05^{\circ}$ is superficially consistent with this picture, understood collectively as a decrease in $U/W$ as the twist angle grows over this range of angles.

For $D<0$, the behavior of the correlated states appears to be less sensitive to twist angle. Correlated states with metallic temperature dependence are observed in devices within the approximate range $1.05^{\circ} \lesssim \theta \lesssim 1.25^{\circ}$. The lone exception is the $\nu=1$ state in device D2, which becomes weakly insulating at temperatures below 1 K. In all devices for which there are suitable measurements, the correlated metallic states at $\nu=1$ and $3$ have an associated AHE. Although overlooked initially, we note that a retrospective analysis of the $\theta=1.05-1.08^{\circ}$ device from Ref.~\cite{Chen2020tMBG} also reveals a similar AHE to the devices we report here. Gradual transitions to insulating behavior are observed upon increasing $B$ at $\nu=1$, whereas abrupt transitions (accompanied by hysteresis in device D2) are observed at $\nu=2$ and $3$. Previous measurements from device D1 in Ref.~\cite{Chen2020tMBG} show that these states are nearly insensitive to $B_{||}$, suggesting that all of the phase transitions are driven by orbital effects.

\begin{table*}[h]
\centering
\begin{tabularx}{0.97\textwidth} { 
  | >{\centering\arraybackslash}X 
  || >{\centering\arraybackslash}X 
  | >{\centering\arraybackslash}X 
  | >{\centering\arraybackslash}X 
  | >{\centering\arraybackslash}X 
  | >{\centering\arraybackslash}X 
  | >{\centering\arraybackslash}X 
  | >{\centering\arraybackslash}X }
\hline
 & $D>0$, $\nu=1$ & $D>0$, $\nu=2$ & $D>0$, $\nu=3$ & $D<0$, $\nu=1$ & $D<0$, $\nu=2$ & $D<0$, $\nu=3$ \\ 
\hhline{|=#=|=|=|=|=|=|}
$0.89^{\circ}$ (Ref.~\cite{Chen2020tMBG}) & Chern insulator (approaching quantization with $C=1$) & weakly insulating & none & none & none & none \\ 
\hline
$0.90^{\circ}$ (Ref.~\cite{Polshyn2020}) & none & none & none & none & none & none \\  
\hline
$1.05-1.08^{\circ}$ (Ref.~\cite{Chen2020tMBG}) & none (weak state emerges in large $B_{||}$) & trivial insulator (gap grows with $B_{||}$, suggesting spin-polarized) & none (weak state emerges in large $B_{||}$) & metallic resistive state with AHE (incipient insulator emerges gradually with $B$) & metallic resistive state (insulator emerges abruptly at finite $B$) & metallic resistive state with AHE (insulator emerges abruptly at finite $B$) \\  
\hline
$1.13^{\circ}$ (device D1) & trivial insulator (AHE upon doping) & trivial insulator & none & metallic resistive state with AHE (insulator emerges gradually with $B$) & metallic resistive state (insulator emerges abruptly at finite $B$) & metallic resistive state with AHE (insulator emerges abruptly at finite $B$) \\  
\hline
$1.14^{\circ}$ (device D3) & trivial insulator (AHE upon doping) & trivial insulator & none & metallic resistive state with AHE (insulator emerges gradually with $B$) & metallic resistive state (insulator emerges abruptly at finite $B$) & metallic resistive state with AHE (insulator emerges abruptly at finite $B$) \\  
\hline
$1.19^{\circ}$ (device D2) & trivial insulator (AHE upon doping) & trivial insulator & obscured by poor contacts & trivial insulator (AHE upon doping) & metallic resistive state (insulator emerges abruptly at finite $B$) & metallic resistive state with AHE (insulator emerges abruptly at finite $B$) \\  
\hline
$1.25^{\circ}$ (Ref.~\cite{Polshyn2020}) & Chern insulator (approaching quantization with $C=2$) & trivial insulator & Chern insulator (approaching quantization with $C=2$) & metallic resistive state & metallic resistive state & metallic resistive state \\  
\hline
$1.385^{\circ}$ (Ref.~\cite{Polshyn2020}) & metallic with weak AHE & trivial insulator & symmetry-broken metal & none & none & none \\ 
\hline
\end{tabularx}
\caption{Summary of the properties of the correlated states in devices D1, D2, and D3, as well as in selected devices from Refs.~\cite{Chen2020tMBG,Polshyn2020}. Entries reading ``none'' indicate that no correlated state is clearly observed or reported.}
\label{tab:summary}
\end{table*}

\clearpage

\begin{figure*}[t]
\includegraphics[width=5.5in]{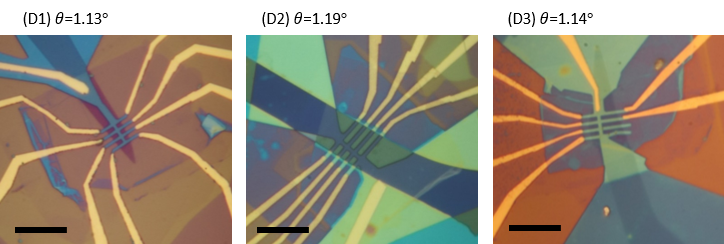} 
\caption{\textbf{Optical microscope images of the three tDMG devices.}
The twist angle of each device is denoted at the top left corner of each image. Devices are encapsulated in BN flakes with thicknesses of 10-30 nm. All scale bars are 10 $\mu$m.
}
\label{fig:deviceimages}
\end{figure*}

\begin{figure*}[t]
\includegraphics[width=6.9in]{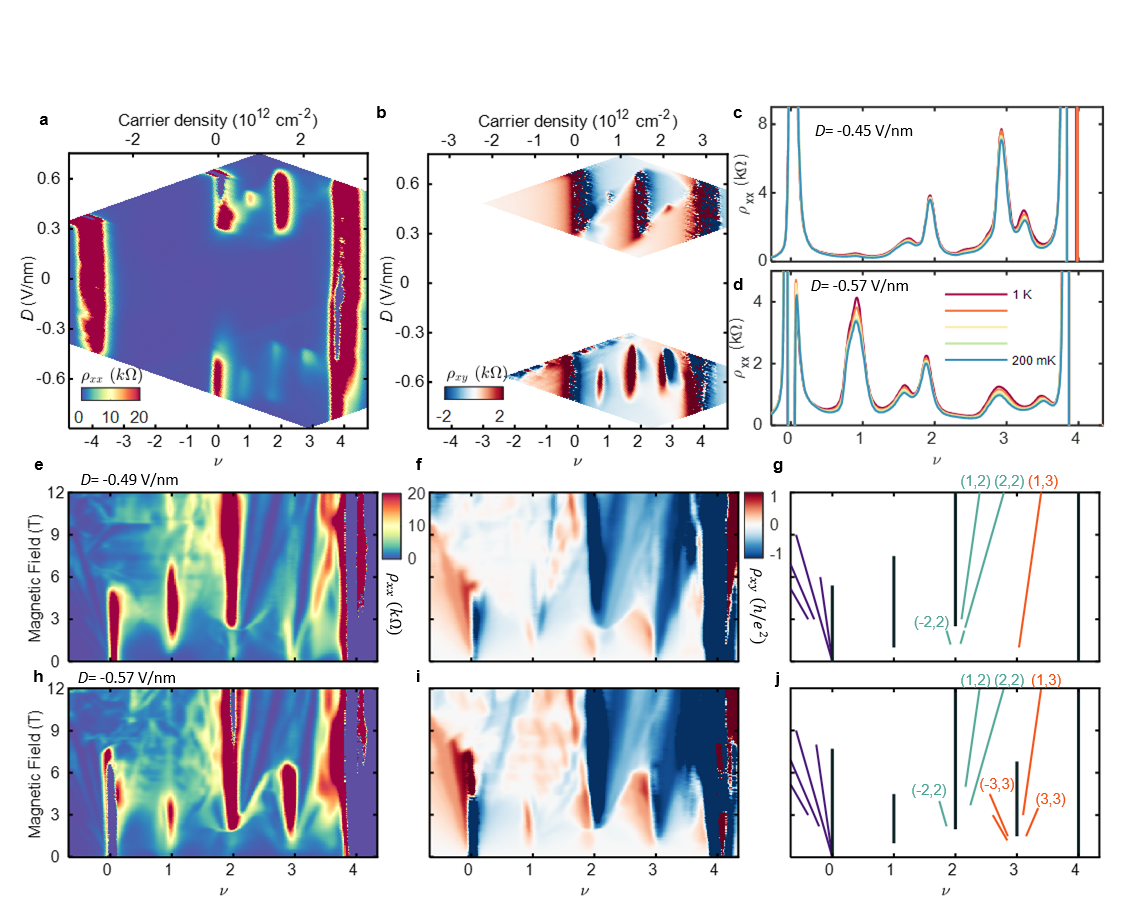} 
\caption{\textbf{Additional transport characterization and Landau fan diagrams for $D<0$ in device D1 ($\theta=1.13^{\circ}$).}
Maps of \textbf{a}, $\rho_{xx}$ at $B=0$ and \textbf{b}, $\rho_{xy}$ antisymmetrized with $|B|=0.5$~T, acquired at $T=300$~mK. The primary difference in $\rho_{xx}$ at $B=0$ compared with the symmetrized map at $|B|=0.5$~T shown in Fig.~\ref{fig:1}a of the main text is the complete absence of a correlated insulating state at $\nu=3$.
\textbf{c-d}, $\rho_{xx}$ as a function of filling factor at $D=-0.45$~V/nm (\textbf{c}) and $D=-0.57$ (\textbf{d}), acquired at different $T$ below 1~K. All correlated states at $\nu=1$, $2$, and $3$ exhibit metallic temperature dependence.
\textbf{e-j}, Landau fan diagrams at $D=-0.49$~V/nm (\textbf{e-g}) and $D=-0.57$~V/nm (\textbf{h-j}), acquired at $T=300$~mK. The leftmost column shows $\rho_{xx}$, the central column shows $\rho_{xy}$, and the rightmost column schematically denotes the observed states following the convention established in Fig.~\ref{fig:3} of the main text. In \textbf{e-g}, we observe a very strong $(1,3)$ state that emerges and becomes quantized at very small $B$. This likely reflects the anticipated $C_v=1$ of the band at $D<0$, however we are unable to determine this unambiguously in the absence of a QAH state at $B=0$. We additionally observe a clear Landau fan corresponding to the correlated state at $\nu=2$ at very low fields in (\textbf{e-g}). These gapped states close at the phase transition ($B \sim 2.8$~T) and immediately reopen at higher field along with a robust trivial insulating state, $(0,2)$. Although this implies a first-order phase transition between two different correlated states, we are not able to unambiguously identify the ground state order of either from our measurements.
}
\label{fig:deviceD1}
\end{figure*}

\begin{figure*}[t]
\includegraphics[width=6.9in]{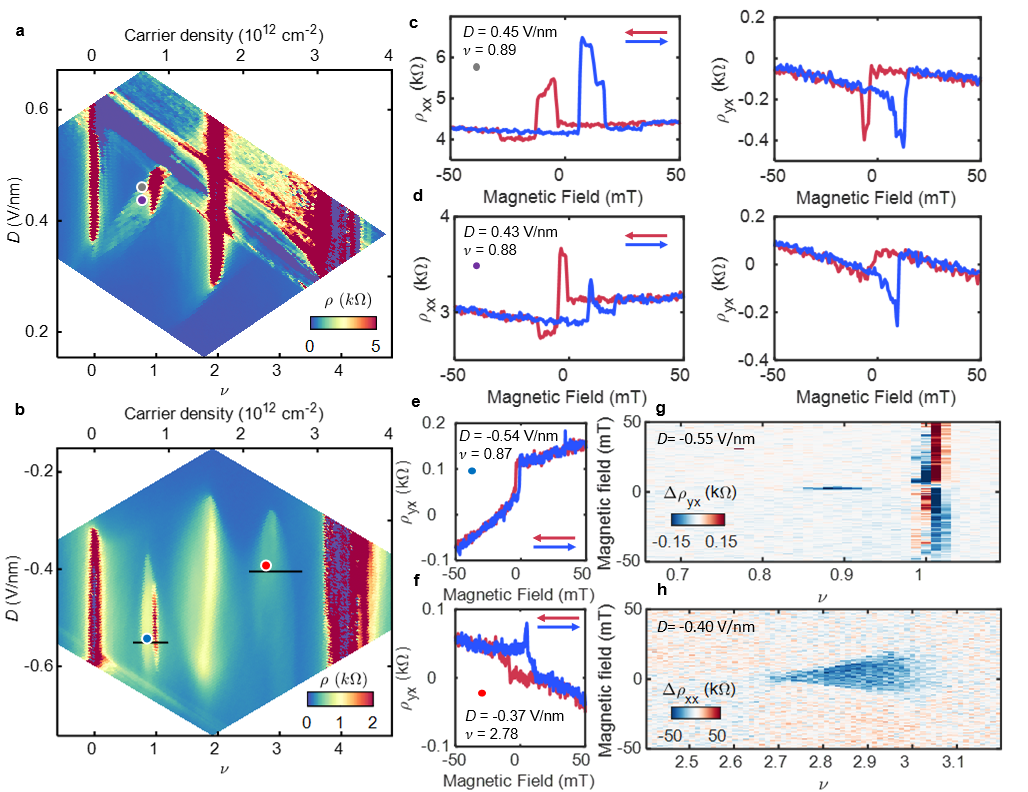} 
\caption{\textbf{Additional transport characterization and AHE in device D2 ($\theta=1.19^{\circ}$).}
Maps of \textbf{a}, $\rho_{xx}$ at $B=0$ and \textbf{b}, $\rho_{xy}$ antisymmetrized with $|B|=0.5$~T, acquired at $T=15$~mK and $T=500$~mK respectively. Artifacts due to poor contacts obscure a portion of the map in (\textbf{a}), preventing an analysis of the $\nu=3$ state for $D>0$ in this device.
\textbf{c-d}, $\rho_{xx}$ (top) and $\rho_{yx}$ (bottom) acquired as $B$ is swept back and forth at $\nu$ and $D$ indicated by the labels and the associated gray and purple markers in (\textbf{a}), acquired at $T=500$~mK. Similar to device D1 shown in Fig.~\ref{fig:1} of the main text, we observe AHE for $D>0$ only within the ``halo'' region ($\nu<1$) associated with the correlated insulating state at $\nu=1$. No obvious AHE signatures are observed precisely at $\nu=1$.
\textbf{e-f}, $\rho_{yx}$ acquired as $B$ is swept back and forth at $\nu$ and $D$ indicated by the labels and the associated blue and red markers in (\textbf{b}). We observe a weak AHE in the correlated metallic states nearby both $\nu=1$ and $3$, qualitatively similar to the behavior of device D1 as shown in Fig.~\ref{fig:1} of the main text. Data are acquired at $T=500$~mK for \textbf{e} and at $T=50$~mK for \textbf{f}.
\textbf{g}, $\Delta \rho_{yx}$ as a function of doping at $D=-0.55$~V/nm. In contrast to all other devices, we observe a trivial insulating state precisely at $\nu=1$, with no clear signatures of an AHE (see also Supplementary Information Figs.~\ref{fig:nu1_absentAHE}c-d).
\textbf{h}, $\Delta \rho_{xx}$ as a function of doping at $D=-0.55$~V/nm. $\Delta \rho_{xx}$ is shown rather than $\Delta \rho_{yx}$ owing to mixing effects observed in the latter for the specific contact pairs tested. Data in \textbf{g-h} is acquired at $T=300$~mK. 
}
\label{fig:deviceD2_AHE}
\end{figure*}

\begin{figure*}[t]
\includegraphics[width=6.9in]{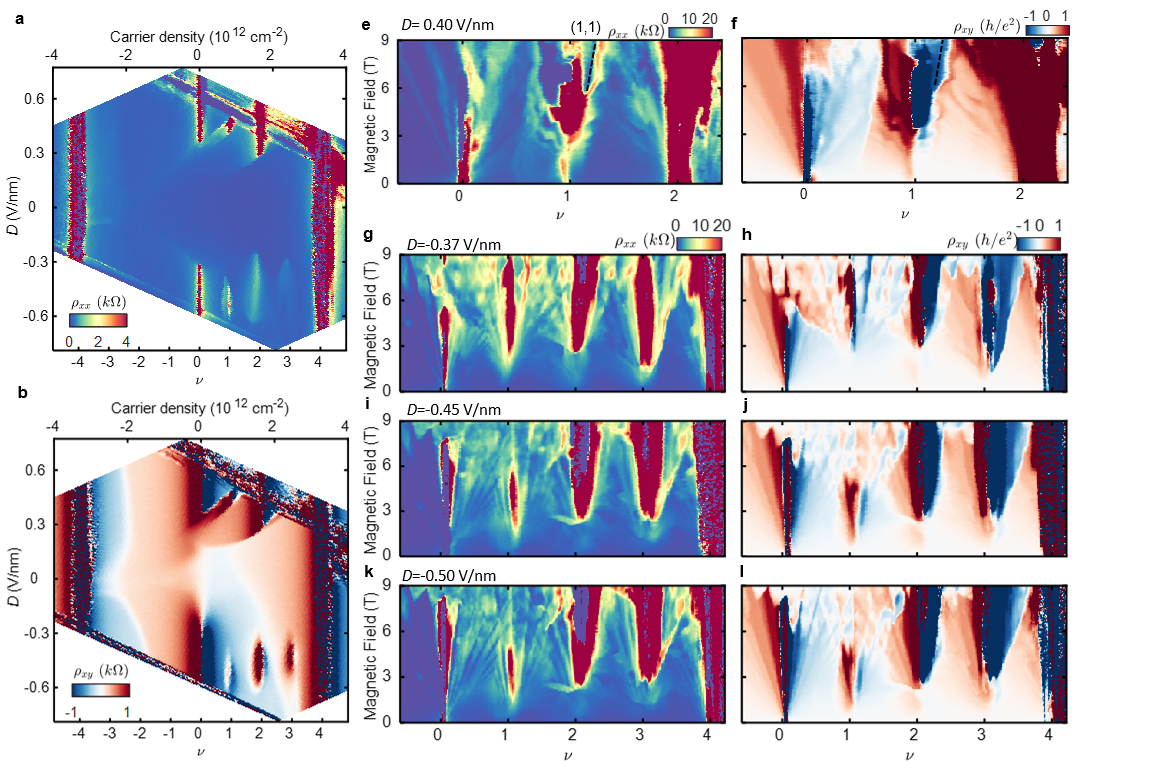} 
\caption{\textbf{Additional transport characterization and Landau fan diagrams in device D2 ($\theta=1.19^{\circ}$).}
Maps of \textbf{a}, $\rho_{xx}$ at $B=0$ and \textbf{b}, $\rho_{xy}$ antisymmetrized with $|B|=0.5$~T, acquired at $T=15$~mK and $T=500$~mK respectively.
\textbf{c-d}, Landau fan diagram at $D=0.40$~V/nm, acquired at $T=100$~mK. Similar to device D1 shown in the main text, we observe a trivial insulating state at $\nu=1$ for small $B$, which appears to cede to a $(1,1)$ Chern insulator state at higher field (marked by dashed black line). States near $\nu=3$ are obscured by poor contacts, and are not shown.    
\textbf{e-j}, Landau fan diagrams at $D=-0.37$~V/nm (\textbf{g-h}), $D=-0.45$~V/nm (\textbf{i-j}), and $D=-0.50$~V/nm (\textbf{k-l}), acquired at $T=100$~mK. The left column shows $\rho_{xx}$ and the right column shows $\rho_{xy}$. We find that the critical fields, $B_{c}$, corresponding to the onset of insulating states at $\nu=1$, $2$, and $3$ depend on $D$. All exhibit signatures of first-order phase transitions at or near $\nu=2$ and $3$, but a continuous onset of insulating behavior at $\nu=1$.
}
\label{fig:deviceD2_fan}
\end{figure*}

\begin{figure*}[t]
\includegraphics[width=6.5in]{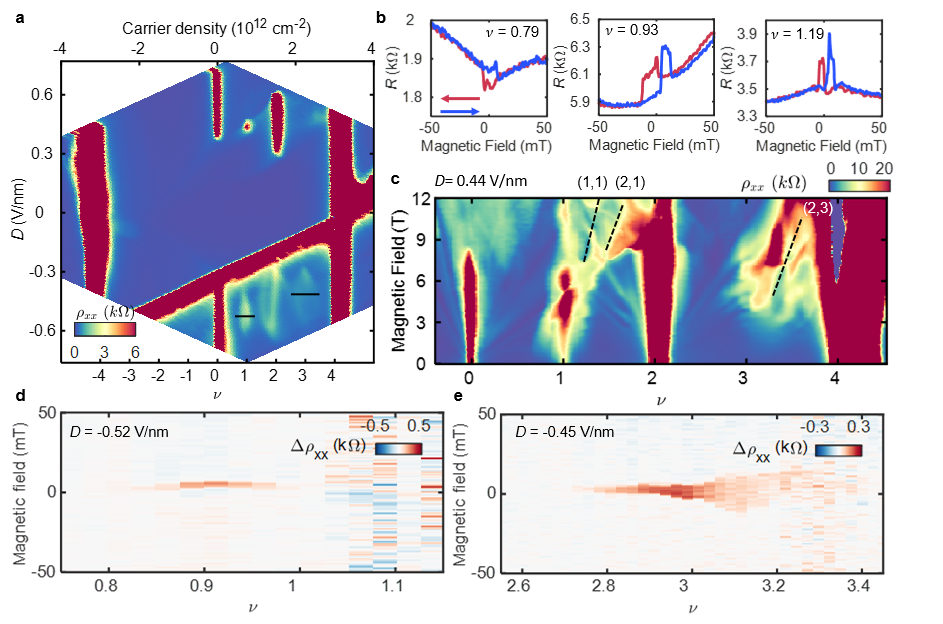} 
\caption{\textbf{Transport characterization, AHE, and Landau fan diagrams in device D3 ($\theta=1.14^{\circ}$).}
\textbf{a}, Map of $\rho_{xx}$ at $B=0$, acquired at $T=15$~mK. This device did not have suitable working contacts to measure the corresponding $\rho_{xy}$ without large mixing with $\rho_{xx}$.
\textbf{b}, AHE shown at different $\nu$ near $\nu=1$ with $D=0.46$~V/nm. $R$ is shown here due to lack of proper $\rho_{yx}$ contacts, it thus exhibits features of both $\rho_{xx}$ and $\rho_{yx}$. Similar to devices D1 and D2, we observe the AHE nearby $\nu=1$, but trivial insulating behavior at $\nu=1$.
\textbf{c}, Landau fan diagram at $D=0.44$~V/nm, acquired at $T=300$~mK. Similar to devices D1 and D2, we observe a trivial insulating state at $\nu=1$ for small $B$, which appears to cede to a sequence of correlated Chern insulator states at higher field. Chern insulator states corresponding to $\nu=3$ emerge only at high field, similar to device D1. 
\textbf{d-e}, $\Delta \rho_{xx}$ as a function of doping at $D=-0.52$~V/nm nearby $\nu=1$ (\textbf{d}) and $D=-0.45$~V/nm nearby $\nu=3$ (\textbf{e}), acquired at $T=300$~mK, as indicated by the black lines in \textbf{a}. We observe clear AHE for a wide range around $\nu=3$, and signatures of a very weak AHE near $\nu \sim 0.9$.
}
\label{fig:deviceD3}
\end{figure*}

\begin{figure*}[t]
\includegraphics[width=5in]{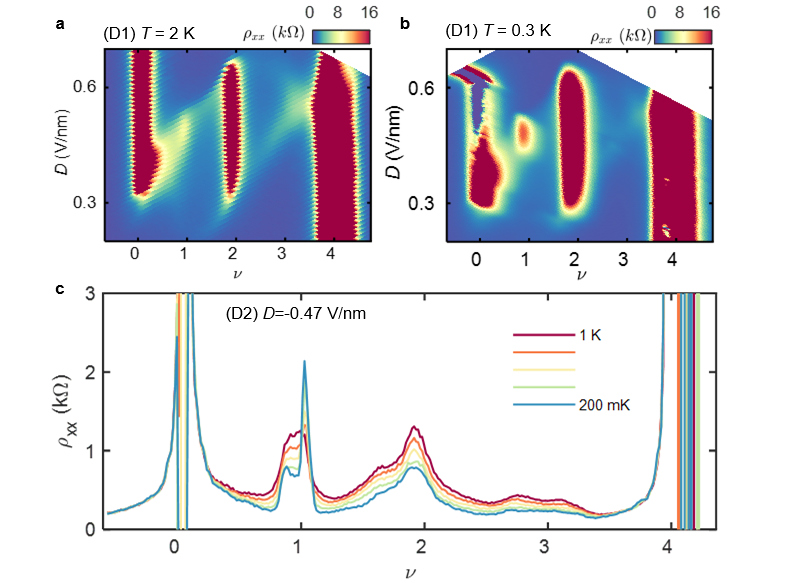} 
\caption{\textbf{Temperature dependence of correlated insulating states at $\nu=1$.}
\textbf{a-b}, Maps of $\rho_{xx}$ for $D>0$ in device D1, acquired at $T=2$~K (\textbf{a}) and $T=0.3$~K (\textbf{b}). Over a small range of $D$, the resistivity at $\nu=1$ increases with decreasing temperature, marking insulating behavior.
\textbf{c}, $\rho_{xx}$ as a function of filling factor at $D=-0.47$~V/nm in device D2, measured at selected temperatures from $1$~K to $200$~mK, in steps of 200 mK. A correlated insulating state emerges precisely at $\nu=1$, as indicated by the sharp peak in $\rho_{xx}$ emerging as the temperature is lowered. All other correlated states remain metallic down to base temperature.
}
\label{fig:nu1_temperature}
\end{figure*}

\begin{figure*}[t]
\includegraphics[width=5in]{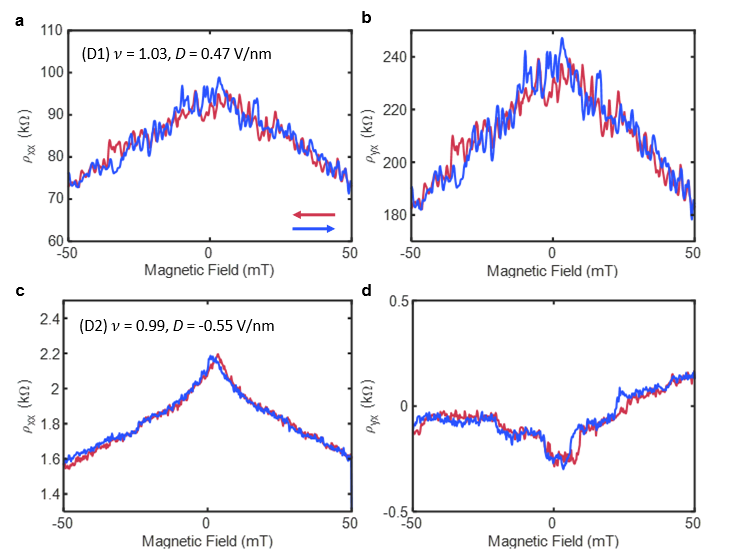} 
\caption{\textbf{Absence of the AHE precisely at $\nu=1$ for insulating states.}
\textbf{a}, $\rho_{xx}$ and \textbf{b}, $\rho_{yx}$ measured as $B$ is swept back and forth at $\nu=1.03$ for $D=0.47$~V/nm in device D1.
\textbf{c}, $\rho_{xx}$ and \textbf{d}, $\rho_{yx}$ measured as $B$ is swept back and forth at $\nu=0.99$ for $D=-0.55$~V/nm in device D2.
In both cases, insulating behavior is observed at $\nu=1$ (see Supplementary Information Fig.~\ref{fig:nu1_temperature}).
Neither exhibit any clear signatures of AHE; most notably, we do not observe a hysteresis loop encircling $B=0$. The state at $\nu=1$ has very high resistance in device D1, and the measured $\rho_{yx}$ appears to have large mixing with $\rho_{xx}$. Owing to the noise in the measurements and the insulating behavior of the state, calculating $\Delta \rho_{yx}$ leads to rapidly oscillating negative and positive values, as seen in Fig.~\ref{fig:1}e of the main text.
We observe abrupt but small jumps at various values of $|B|>0$ in (\textbf{d}). We do not know the origin of these features, and although they may potentially indicate some form of disordered magnetism, they are distinct from the single hysteresis loops around $B=0$ observed for $\nu<1$, as shown in Supplementary Information Figs.~\ref{fig:deviceD2_AHE}e and g. 
}
\label{fig:nu1_absentAHE}
\end{figure*}

\begin{figure*}[t]
\includegraphics[width=6in]{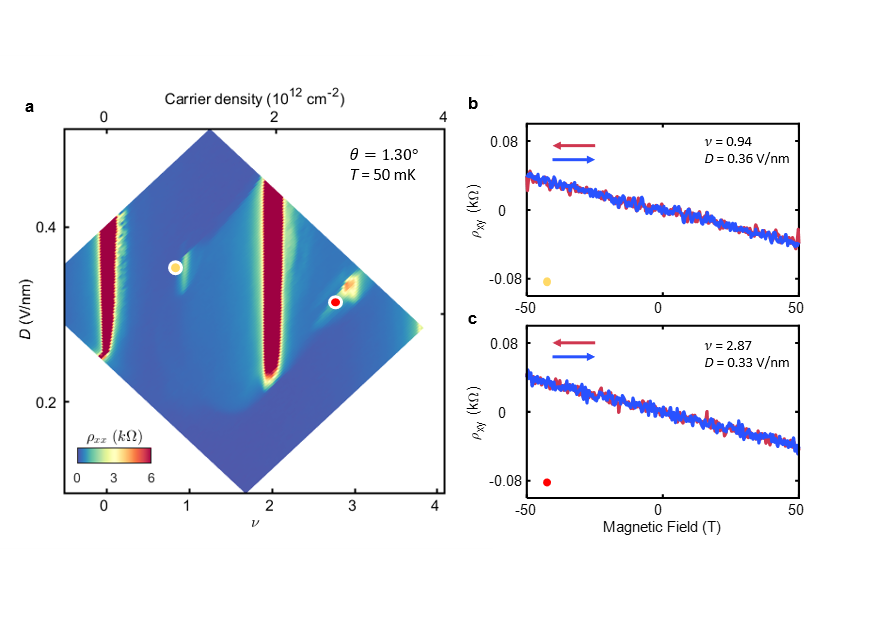} 
\caption{\textbf{Absence of the AHE near $\nu=1$ and $3$ in tDBG.}
\textbf{a}, Map of $\rho_{xx}$ for a twisted double bilayer graphene (tDBG) device with $\theta=1.30^{\circ}$. We observe a robust correlated insulating state at $\nu=2$, and very weakly developed correlated insulating states at $\nu=1$ and $3$. We also observe small ``halo'' features surrounding $\nu=1$ and $3$, indicating the formation of new symmetry-broken Fermi surfaces at each.
\textbf{b-c}, Antisymmetrized $\rho_{xy}$ measured as the field is swept back and forth near $\nu=1$ (\textbf{b}) and $\nu=3$ (\textbf{c}), at $\nu$ and $D$ indicated by the labels and the associated yellow and red markers in (\textbf{a}). We do not observe any signatures of an AHE or hysteresis in these measurements, nor at any other measured value of $\nu$ and $D$.
}
\label{fig:tDBG_noAHE}
\end{figure*}

\end{document}